\documentclass[aoas]{imsart}

\RequirePackage{amsthm,amsmath,amsfonts,amssymb}
\RequirePackage[authoryear]{natbib}
\RequirePackage[colorlinks,citecolor=blue,urlcolor=blue]{hyperref}
\RequirePackage{graphicx}
\usepackage{dsfont}
\usepackage{booktabs}
\usepackage{float}
\usepackage{threeparttable}
\usepackage{enumitem}

\startlocaldefs
\theoremstyle{plain}

\newtheorem{theorem}{Theorem}[section]
\newtheorem{lemma}[theorem]{Lemma}

{
  \theoremstyle{plain}
  \newtheorem{assumption}{}
}

\theoremstyle{definition}
\newtheorem{definition}{Definition}

\newtheorem{algorithm}{Algorithm}
\newtheorem{remark}{Remark}
\newcommand{\norm}[1]{\|#1\| }
\DeclareMathOperator*{\argmin}{arg\,min}
\renewcommand{\P}{\mathsf{P}}
\newcommand{\E}{\mathds{E}}

\renewcommand{\d}{\mathsf{d}}

\newcommand{\m}{\mathsf{m}}
\newcommand{\rr}{\mathsf{r}}
 
\newcommand{\g}{\mathsf{g}}

\newcommand{\bb}{\mathsf{b}}

\newcommand{\one}{\mathds{1}}

\newcommand{\Vone}{\mathbf{V_1}}
\newcommand{\vone}{\mathbf{v_1}}

\newcommand{\W}{\mathbf{W}}
\newcommand{\w}{\mathbf{w}}

\DeclareMathOperator*{\argmax}{arg\,max}
\newcommand{\indep}{\perp\!\!\!\perp}
\newcommand{\CPE}{\Tilde{\tau}(\vone, v_2)}
\newcommand{\CPEa}{\tilde{f}(a, \vone,v_2)}
\newcommand{\CPEone}{\tilde{f}(1, \vone,v_2)}
\newcommand{\CPEzero}{\tilde{f}(0, \vone,v_2)}
\newcommand{\EIF}{\xi(\mathbf{O}; \eta)}
\newcommand{\EIFa}{\xi(\mathbf{O},a,v_2; \eta)}

\endlocaldefs

\begin{document}

\begin{frontmatter}
\title{Nonparametric estimation of an optimal treatment rule with fused randomized trials and missing effect modifiers}
\runtitle{Estimating the ODTR in fused trials with missing modifiers}

\begin{aug}
\author[A]{\fnms{Nicholas}~\snm{Williams}\ead[label=e1]{ntw2117@cumc.columbia.edu}},
\author[A]{\fnms{Kara}~\snm{Rudolph}\ead[label=e2]{kr2854@cumc.columbia.edu}}
\and
\author[B]{\fnms{Iván}~\snm{Díaz}\ead[label=e3]{ivan.diaz@nyulangone.org}}
\address[A]{Department of Epidemiology, Columbia University \printead[presep={ ,\ }]{e1,e2}}

\address[B]{Division of Biostatistics, New York University \printead[presep={,\ }]{e3}}
\end{aug}

\begin{abstract}
    A fundamental principle of clinical medicine is that a treatment should only be administered to those patients who would benefit from it. Treatment strategies that assign treatment to patients as a function of their individual characteristics are known as dynamic treatment rules. The dynamic treatment rule that optimizes the outcome in the population is called the optimal dynamic treatment rule. Randomized clinical trials are considered the gold standard for estimating the marginal causal effect of a treatment on an outcome; they are often not powered to detect heterogeneous treatment effects, and thus, may rarely inform more personalized treatment decisions. The availability of multiple trials studying a common set of treatments presents an opportunity for combining data, often called data-fusion, to better estimate dynamic treatment rules. However, there may be a mismatch in the set of patient covariates measured across trials. We address this problem here; we propose a nonparametric estimator for the optimal dynamic treatment rule that leverages information across the set of randomized trials. We apply the estimator to fused randomized trials of medications for the treatment of opioid use disorder to estimate a treatment rule that would match patient subgroups with the medication that would minimize risk of return to regular opioid use.
\end{abstract}


\end{frontmatter}

\section{Introduction} The clinical utility of a treatment refers to the probability that the treatment will improve patient outcomes. Accordingly, the goal of personalized medicine is to maximize clinical utility by tailoring interventions to patients based on their clinical characteristics \citep{kosorok2019precision,kosorok2021introduction}. In epidemiological jargon, treatment decisions based on patient characteristics---specifically, characteristics that modify the effect of treatment on an outcome (i.e., effect modifiers)---are referred to as dynamic treatment rules (DTRs), and the rule that maximizes the benefit of treatment in the population is known as the optimal dynamic treatment rule (ODTR) \citep{murphy2003optimal}.

Randomized clinical trials (RCTs) are considered the gold standard for evaluating treatment efficacy \citep{imbens2015causal}. However, these results may not generalize to all subpopulations or account for subgroup heterogeneity in treatment response \citep{stuart2015assessing}. As such, secondary analyses of RCTs offer an opportunity to learn ODTRs and improve clinical decision-making. These analyses are particularly valuable because treatment randomization ensures the no-unmeasured-confounding assumption required for drawing causal conclusions \citep{imbens2015causal}, and because RCTs typically collect high-quality data.

Despite these benefits, RCTs powered to detect marginal effects typically have insufficient sample sizes to precisely estimate conditional treatment effects \citep{kent2010assessing}. One solution is to pool data from multiple RCTs---often referred to as data fusion---into a single harmonized dataset \citep{brantner2023methods,brantner2024comparison}. Unfortunately, differences in trial protocols may result in systemic missingness of potentially important effect modifiers. We address this challenge here by proposing a estimator for the ODTR that leverages information from all trials in the pooled dataset based on a novel decomposition of the conditional average treatment effect (CATE). Our estimator is non-parametric, doubly-robust to nuisance parameter model misspecification, and capable of leveraging flexible machine learning algorithms while retaining statistical guarantees, under conditions. 

\subsection{Motivating application}

Opioid use disorder (OUD) is a leading cause of death in the United States \citep{cdc2024report}. The standard of care for OUD is medication assisted treatment (MAT) with one of three medications: buprenorphine-naloxone (BUP-NX), extended release injection naltrexone (XR-NTX), or methadone \citep{volkow2019prevention,williams2018developing}. The medication a patient receives generally is a function of where they receive treatment---traditional opioid treatment programs (i.e., methadone clinics) or office-based clinics---which itself is influenced by the demographic characteristics of their community \citep{goedel2020association,hansen2013variation,hansen2016buprenorphine}. This one-size-fits-all approach for treatment based on geographic location is likely unideal \citep{rudolph2021optimizing,rudolph2023optimally,williams2024learning} given that the population of individuals with OUD is heterogeneous. 

Recent work has focused on learning dynamic treatment rules for assigning BUP-NX or XR-NTX (the two medications offered by office-based clinics) based on patient characteristics to minimize the risk of relapse.\citep{rudolph2021optimizing,rudolph2023optimally} This research used data from comparative effectiveness randomized trials for OUD treatment that were part of the National Institute on Drug Abuse Clinical Trials Network (CTN). Using data from CTN0051 \citep{lee2018comparative}, \cite{rudolph2021optimizing} learned a DTR for assigning patients with OUD either BUP-NX or XR-NTX to minimize relapse. They found that a treatment rule based on patient characteristics would reduce risk of relapse by 24-weeks by 12\% compared to randomly assigning treatment. Importantly, \cite{rudolph2021optimizing} also found that the most important effect modifier was whether or not a patient had stable housing. The DTR from \cite{rudolph2021optimizing} is clinically relevant because BUP-NX is a take-home prescription medication that patients need to keep track of and self-administer daily, which may be a significant challenge for patients with unstable housing. XR-NTX is a single injection administered every 4-weeks. However, treating OUD with XR-NTX typically requires the patient be completely withdrawn from opioids to avoid precipitated withdrawal \citep{jarvis2018extended}, which creates a significant barrier to induction. In a follow-up analysis, \cite{rudolph2023optimally} learned a DTR for assigning BUP-NX or XR-NTX that would minimize relapse by 12-weeks of treatment using a harmonized cohort of data from CTN0051 and CTN0030 \citep{weiss2011adjunctive}. Housing status, however, was not recorded in CTN0030 and was excluded as an effect modifier from their analysis. Our goal is to improve upon the ODTR learned by \cite{rudolph2023optimally} by learning a rule that incorporates housing status in the pooled data.

\section{Estimating the ODTR with missing modifiers}

\subsection{Notation}

Let $\mathbf{O} = (S, \mathbf{W},\mathbf{V_1}, SV_2, A, Y)\sim P$ represent the observed data sampled from an unknown probability distribution $P$; $A$ is a binary treatment, $Y$ a binary outcome, $S$ an indicator of trial membership, $\W$ a vector of covariates (needed for confounding adjustment and/or efficiency gains), $\Vone$ a vector of, possibly high-dimensional, observed effect modifiers of $A \rightarrow Y$, and $V_2$ a discrete effect modifier of $A \rightarrow Y$ that is only measured for observations where $S=1$. Assume we sample $\mathbf{O}_1, ..., \mathbf{O}_n$ i.i.d observations of $\mathbf{O}$. We let $\norm{f} = \int\{f(\mathbf{o)}^2\, d\P(\mathbf{o})\}^{-1/2}$ denote the $L_2$-norm and $\E[f] = \int f(\mathbf{o)}\,d\P(\mathbf{o})$ denote the expectation of $f(\mathbf{O})$ with respect to $\P$. Capital letters (e.g., $A$) indicate random variables while lower-case letters (e.g., $a$) indicate realizations of a random variable; $\mathcal{X}$ represents the support of $X$. We use $Y_a$ to denote the counterfactual value of $Y$ that would be observed if $A=a$ were assigned. 

\begin{figure}[H]
  \centering
    \includegraphics[width=0.8\textwidth]{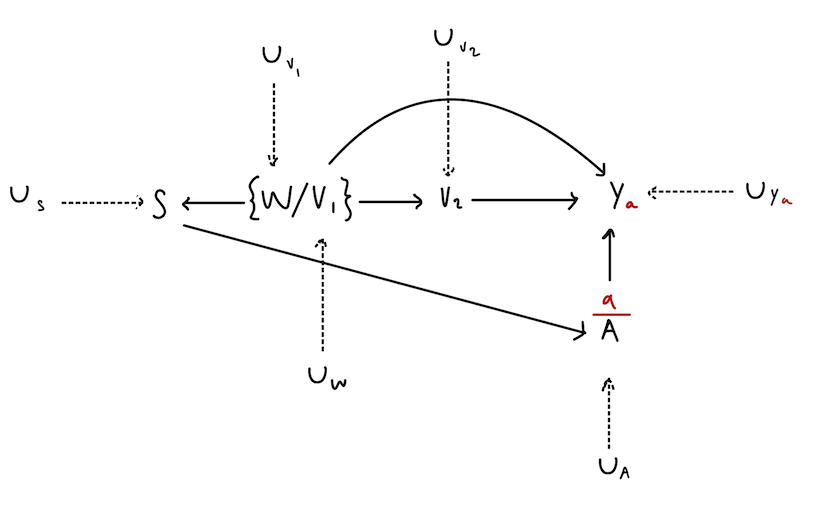}
    \caption{Single world intervention graph}
    \label{fig:dag2}
\end{figure}

We formalize the definition of causal effects using nonparametric structural equation models (NPSEM) \citep{pearl2009causality}. Assume the existence of deterministic functions $\W = f_\W(U_\W)$, $\Vone = f_{\Vone}(U_{\Vone})$, $S = f_S(\W, \Vone, U_S)$, $V_2 = f_{V_2}(\W, \Vone, U_{V_2})$, $A = f_A(S, U_A)$, and $Y = f_{Y}(A, \W, \Vone, V_2, U_Y)$, where variables $U = (U_S, U_\W,U_{\Vone}, U_{V_2}, U_A, U_Y)$ represent unmeasured mutually independent variables. A single world intervention graphs (SWIG) representing the assumed NPSEM is shown in Figure \ref{fig:dag2} \citep{richardson2013single}. Notably, there is an absence of an arrow from $V_2$ to $S$. We also allow an arrow from $S$ to $A$ in our causal model. Practically, this can occur when the treatment arms are not the same across studies---as is the case in our motivating example. However, our identification result still applies if treatment remains randomized in the pooled data. 

\subsection{Finding a target parameter}

In this section, we turn our attention to finding a statistical estimand that leverages data from all available trials and can be used to estimate the ODTR. 

\begin{definition}
The ODTR is the decision rule function $\d(\mathbf{v})$ in the set of all possible decision rules $\mathcal{D}$ that takes as input effect modifiers, $\mathbf{v}$, and outputs a treatment decision as
\[\d_\text{opt}(\mathbf{v})=\argmax_{\d\in\mathcal{D}}\E[Y_\d].\]
where $Y_\d$ is the counterfactual value of $Y$ that would be observed if treatment were assigned according to $\d$.
\end{definition}

There are two general approaches to learning the ODTR: 1) those based on a regression of the conditional average treatment effect (CATE), including Q-learning \citep{watkins1989learning,moodie2012q,williams2024learning} and A-learning \citep{murphy2003optimal}; and 2) direct classification methods like outcome-weighted learning \citep{zhao2012estimating}, residual-weighted learning \citep{zhou2017residual}, and augmented outcome-weighted learning \citep{liu2018augmented}. In this paper, we focus on estimation based on the CATE. Let 
\begin{equation}\label{eq:psia}
       f(a, \vone,v_2) = \P(Y_a=1\mid \Vone=\vone,V_2=v_2).
\end{equation} 
Then the CATE is defined as
\begin{equation}\label{eq:CATE}
       \tau(\vone,v_2) = f(1, \vone,v_2) - f(0, \vone,v_2).
\end{equation}
Assuming $\P(\tau(\vone,v_2) = 0) = 0$, the ODTR can be defined as the sign of the CATE
\begin{equation}\label{eq:odtr}
    \d_\text{opt}(\mathbf{v}) \equiv \one(\tau(\mathbf{v_1},v_2) > 0).
\end{equation}

Because we only observe $V_2$ in trial $S = 1$ our goal is to express Equation \ref{eq:psia} in a way that it is identifiable with the available data from the pooled trials. Specifically, we can rewrite Equation \ref{eq:psia} as
\begin{equation}\label{eq:newpsia}
    f(a, \vone,v_2) = \frac{\P(V_2=v_2\mid Y_a=1, \Vone=\vone)\P(Y_a=1\mid \Vone=\vone)}{\P(V_2=v_2\mid \Vone=\vone)}, 
\end{equation}
where we can identify $\P(V_2=v_2\mid Y_a=1, \Vone=\vone)$ and $\P(V_2=v_2\mid \Vone=\vone)$ from the subset of observations where $S = 1$, and $\P(Y_a=1\mid \Vone=\vone)$ using the entire pooled data.
Notice, however, that when the CATE is expressed as a contrast of $f(1, \vone, v_2)$ and $f(0, \vone, v_2)$ (expressed as Equation \ref{eq:newpsia}), its sign depends only on the numerator, since only the numerator is a function of $Y_a$. Denote $\CPEa = \P(V_2=v_2\mid Y_a=1, \Vone)\P(Y_a=1\mid \Vone=\vone)$. We can then define the numerator of $\tau(\vone,v_2)$ as
\begin{equation}
    \tilde{\tau}(\vone, v_2) = \CPEone - \CPEzero.
\end{equation}

\begin{remark}\label{lem:odtr}
$\one(\Tilde{\tau}(\vone, v_2) > 0) = 1\iff \d_\text{opt}(\mathbf{v}) = 1$.
\end{remark}
Remark \ref{lem:odtr} indicates that to learn the ODTR we do not need to estimate the full CATE; we only need to estimate $\CPE$. For this reason, we can consider $\CPE$ as a proxy for the additive effect of $A$ on $Y$ within the strata $\{\Vone=\vone, V_2=v_2\}$ and refer to $\CPE$ as the conditional proxy effect (CPE) henceforth. This finding is particularly useful because the alternative expression for the CATE is a ratio. Even if the numerator and denominator are similarly structured, the ratio may be more complex (e.g., non-smooth), or highly-variable if the denominator is close to zero.

\subsection{Identification}\label{sec:identification}

As discussed in the previous section, our target parameter is the conditional proxy effect, $\CPE = \CPEone - \CPEzero$. 
To proceed, we must find an expression for $\CPEa$ that can be written in terms of the observed data. We make the following set of assumptions to identify the CPE.

\begin{assumption}\label{ass:consistency}
    Consistency: $A = a \implies Y = Y_a$.
\end{assumption}

\begin{assumption}\label{ass:pos}
Positivity: $\P(A = a\mid \W,\mathbf{V_1}) > \epsilon > 0\, \forall\, (\W\times\Vone)\in (\mathbf{\mathcal{W}}\times \mathbf{\mathcal{V}_1})$. 
\end{assumption}

\begin{assumption}\label{ass:noconfounding}
Exchangeability: $Y_a \indep A\mid \mathbf{W},\Vone$.
\end{assumption}

\begin{assumption}\label{ass:amod}
    Unmeasured modifier treatment exchangeability: $V_2\indep A\mid \mathbf{W}, \Vone, 
    Y_a$. 
\end{assumption}

\begin{assumption}\label{ass:smod}
Unmeasured modifier $S$-admissibility: $V_2\indep S\mid \W,\Vone,Y_a$. 
\end{assumption}

\begin{assumption}\label{ass:joint}
Joint positivity:  $\P(Y=1, A=a,S=1\mid \W,\Vone)>\epsilon > 0\, \forall\,(\W\times\Vone)\in (\mathcal{W},\mathcal{V}_1)$.
\end{assumption}

Assumptions \ref{ass:consistency}, \ref{ass:pos}, \ref{ass:noconfounding} are standard for identification in the causal inference literature. Assumption \ref{ass:amod} states that conditional on $\W$, $\Vone$, and the potential outcomes $Y_a$ the missing modifier and treatment are independent. Assumption \ref{ass:smod} states that conditional on $\W$, $\Vone$, and the potential outcomes $Y_a$ the probability of $V_2 = v_2$ is the same across studies; it is similar to the standard $S$-admissibility assumption from the transport literature \citep{pearl2011transportability}. We discuss these assumptions in the context of our application in \S\ref{sec:implications}.

\begin{lemma}\label{th:identification}
    Under assumptions \ref{ass:consistency}-\ref{ass:joint}, $\CPEa$ is identified from the observed data as
        \begin{equation}
        \int \bb(a,v_2,\vone)\m(a,\vone)\P(\mathbf{w}\mid\vone)\,d\mathbf{w},
    \end{equation}
    where 
    \begin{align*}
    \bb(a,v_2,\vone) &=\P(V_2=v_2\mid Y=1,A=a,\Vone=\vone,\W,S=1) \\
    \m(a,\vone) &= \E[Y\mid A=a,\Vone=\vone,\W].
\end{align*}
\end{lemma}

\subsection{Estimation}

A variety of options exists for estimating $\CPE$. Perhaps the simplest approach, we could use a naïve plugin estimator of the CPE by regressing $\hat{\bb}(1, v_2,\vone)\hat{\m}_1(\vone) - \hat{\bb}(0, v_2,\vone)\hat{\m}_0(\vone)$ on $\Vone$, where $\hat{\bb}(a, v_2,\vone)$ and $\hat{\m}_a(\vone)$ are estimates of ${\bb}(a, v_2,\vone)$ and ${\m}_a(\vone)$. Alternatively, if $\vone$ is low-dimensional, the CPE could instead be estimated with a one-step estimator by deriving the efficient influence function of $\CPE$ and---for each value $\vone \in \mathcal{V}_1$---solving the corresponding estimating equation where the EIF is set equal to zero. In practice, though, we may not be willing to restrict the ODTR to low-dimensional and discrete covariates. If $\Vone$ is continuous, $\CPE$ is not pathwise differentiable and the previous estimator would no longer be feasible. 

Instead, we propose a cross-fit nonparametric estimator using a doubly-robust unbiased transformation pseudo-regression. That is, we propose to find a doubly-robust mapping $\EIFa$ of the observed data $\mathbf{O}$ and nuisance parameters $\eta$ such that 
\[\E[\EIFa\mid \Vone=\vone] = \tilde f(a,\vone,v_2).\]
The transformation $\EIFa$ is doubly-robust in the sense that the above equality will still hold even if some values of $\eta$ are incorrect. To find such a function, we first note that 

$\E[\EIF] = \kappa(a,v_2)$ where 
\begin{align*}
    \kappa(a,v_2) = \iint \bb(a,v_2,\vone)\m(a,\vone)\P(\mathbf{w}\mid\vone)\P(\vone)\,d\mathbf{w}\,d\vone.
\end{align*}
The natural choice is then to use $\EIFa$ as the uncentered efficient influence function of $\kappa(a,v_2)$. This strategy is frequently referred to as the DR-Learner \citep{van2006statistical,kennedy2023towards}, and has been used for flexible estimation of the CATE \citep{van2006statistical, van2015targeted, luedtke2016super, williams2024learning} and the causal dose-response curve of a continuous treatment \citep{kennedy2017non}. Cross-fitting is a sample splitting technique similar to cross-validation that yields desirable statistical guarantees \citep{zivich2021machine} and may remedy the problem of random seed dependence commonly observed when using data-adaptive machine learning \citep{schader2024don, williams2025seeds}. 

It will be helpful to define the additional nuisance parameters 
\begin{align*}
    \g(a,\vone) &= \P(A=a\mid\Vone=\vone,\W)\\
    \rr(a,\vone) &= \P(Y=1,A=a,S=1\mid \Vone=\vone,\W),
\end{align*}
and the weights
\[
\begin{alignedat}{2}
    \alpha_\g(a,\vone) &=\frac{\one(A=a)}{\g(a,\vone)} 
    &\quad \alpha_{\rr}(a,\vone) &= \frac{\one(A=a,Y=1,S=1)}{\rr(a,\vone)}.
\end{alignedat}
\] 
Moreover, let
\begin{align}
    \varphi_\bb(a,v_2) &= \alpha_\rr(a,\Vone)\big\{\one(V_2=v_2)\m(a,\Vone)- \bb(a,v_2,\Vone)\m(a,\Vone)\big\} \\
    \varphi_\m(a,v_2) &= \alpha_\g(a,\Vone)\big\{Y\bb(a,v_2,\Vone) - \bb(a,v_2,\Vone)\m(a,\Vone)\big\}.
\end{align}
The uncentered efficient influence function of $\kappa(a,v_2)$ is then defined as
\begin{equation}
\xi(\mathbf{o},a,v_2) =\varphi_\bb(a,v_2) + \varphi_\m(a,v_2) + \bb(a,v_2,\Vone)\m(a,\Vone).
\end{equation}
The proof is provided in the supplementary materials. Additionally, we let $\eta$ denote $(\g, \m, \rr, \bb)$ and $\eta' = (\g', \m', \rr', \bb')$ denote some value of $\eta$. 

\begin{lemma}\label{th:remainder}
    Let
    \begin{align*}
        C'_\bb &= \m'(a,\vone)\frac{\big(\rr(a,\vone) - \rr'(a,\vone)\big)}{\rr'(a,\vone)}\big\{\bb(a,v_2,\vone) - \bb'(a,v_2,\vone)\big\}\\
        C'_\m &= \bb'(a,v_2,\vone)\frac{\big(\g(a,\vone) - \g'(a,\vone)\big)}{\g'(a,\vone)}\big\{\m(a,\vone) - \m'(a,\vone) \big\}\\
        C'_\kappa &= \big(\bb(a,v_2,\vone) - \bb'(a,v_2,\vone)\big)\big(\m(a,\vone) - \m'(a,\vone)\big)\text{.}
    \end{align*}
    Define the second order remainder term
    \begin{equation}
    \text{Rem}(a, \vone, v_2;\eta') = \E[-C'_\bb - C'_\m + C'_\kappa\mid \Vone=\vone]\text{.}
    \end{equation}
    Then, 
    \begin{equation}
        \tilde f(a,\vone,v_2) = \E[\xi(\mathbf{O}, a, v_2;\eta')\mid \Vone=\vone] + \text{Rem}(a, \vone, v_2;\eta').
    \end{equation}
    If $\eta'$ is defined such that (i) $\g' = \g$ and $\bb' = \bb$, (ii) or $\rr' = \rr$ and $\m' = \m$, (iii) or $\bb' = \bb$ and $\m' = \m$ then $\E[\xi(\mathbf{O}, a, v_2;\eta')\mid \Vone=\vone] =  \tilde f(a,\vone,v_2)$.
\end{lemma}
Lemma \ref{th:remainder} is similar to a von Mises expansion \citep{mises1947asymptotic,robins2009quadratic} and exposes why this transformation is robust to model misspecification---if $\g' = \g$ and $\bb' = \bb$, or $\rr' = \rr$ and $\m' = \m$, or $\bb'=\bb$ and $\m' = \m$ then the second order remainder will be zero. The proof of Lemma \ref{th:remainder} is provided in the supplementary materials. Estimation of $\CPE$ can be performed as follows.

\begin{algorithm}[DR-Learner with cross-fitting]\label{alg:drlearner}
Set $V_2 = v_2'$. Let ${\cal D}_1, \ldots, {\cal D}_J$ denote a random partition of data with indices $i \in \{1, \ldots, n\}$ into $J$ prediction sets of approximately the same size such that 
$\bigcup_{j=1}^J {\cal D}_j = \{1, \ldots, n\}$. For each $j$, the training sample is given by ${\cal T}_j = \{1, \ldots, n\} \setminus {\cal D}_j$. Let $\hat f_{j}$ denote the estimator of $f$, obtained by training the corresponding prediction algorithm using only data in the sample ${\cal T}_j$.
    \begin{enumerate}[itemsep=0.5em, parsep=0.25em]
        \item Estimate the nuisance parameters $\hat\eta_j$. These regressions can be performed using any appropriate off-the-shelf method in the statistics and machine learning literature. For each $j \in \{1, ..., J\}$ and $a \in \{0,1\}$:
        \begin{enumerate}
            \item Construct estimates $\hat\g_{j}(a, \Vone)$ of $\g(a,\Vone)$ by training with data ${\cal T}_j$ and predicting on data ${\cal D}_j$.
            \item Construct estimates $\hat\m_{j}(a,\Vone)$ of $\m(a,\Vone)$ by training with data ${\cal T}_j$  and predicting on data ${\cal D}_j$.
            \item Construct estimates $\hat\bb_{j}(a, v_2', \Vone)$ of $\bb(a, v_2', \Vone)$ using the subset of ${\cal T}_j$ where $Y = 1$ and $S = 1$ for training and predicting on data ${\cal D}_j$.
            \item Construct estimates $\hat\rr_{j}(a,\Vone)$ of $\rr(a,\Vone)$ by training with data ${\cal T}_j$ and predicting on data ${\cal D}_j$. This can be performed by expressing $\hat\rr(a,\Vone) = \hat\P(Y=1\mid A=a,S=1, \Vone,\W)\hat\P(A=a \mid S=1, \Vone,\W)\hat\P(S=1 \mid \Vone,\W)$ and estimating each component. 
        \end{enumerate}
        \item Estimate the second-stage pseudo-outcome regressions. Using $\hat\eta = \bigcup_{j=1}^J \hat\eta_j$, for each $a \in \{0,1\}$ construct $\hat\xi(\mathbf{O}, a,v_2';\hat\eta)$ and regress it on covariates $\Vone$---again using any off-the-shelf model---which yields the prediction functions
        \[\hat{\tilde f}(a, \vone,v_2') = \hat\E[\hat\xi(\mathbf{O}, a,v_2';\hat\eta) \mid \Vone = \vone].\]
        \item The final estimate of $\hat{\tilde\tau}(\vone,v_2')$ is given by $\hat{\tilde f}(1,\vone,v_2') - \hat{\tilde f}(0,\vone,v_2')$.
    \end{enumerate}
\end{algorithm}

\begin{lemma}\label{th:rule1}Assume that one of
\begin{enumerate}
    \item $\norm{\hat\bb(a, v_2,\vone) - \bb(a,v_2,\vone)} = o_\P(1)$ and $\norm{\hat\m(a,\vone)-\m(a,\vone)} = o_\P(1)$
    \item $\norm{\hat\g(a,\vone) - \g(a,\vone)} = o_\P(1)$ and $\norm{\hat\m(a,\vone)-\m(a,\vone)} = o_\P(1)$
    \item $\norm{\hat\rr(a,\vone)-\rr(a,\vone)} = o_\P(1)$ and $\norm{\hat\bb(a, v_2,\vone) - \bb(a,v_2,\vone)} = o_\P(1)$
\end{enumerate}
holds. Then, $\text{Rem}(a, \vone, v_2;\eta') = o_\P(1)$.
\end{lemma}

To understand the importance of Lemma \ref{th:rule1}, note---in a slight abuse of notation---that $\norm{\hat{\tilde f} - \tilde f} \leq \norm{\hat{\tilde f} - \tilde f^*} + \norm{\tilde f^* - \tilde f}$, where $\tilde f^*$ is the, so-called, oracle estimator obtained by regressing $\EIFa$---constructed using the true values of $\eta$---on $\Vone$ \citep{kennedy2017non}. This implies that if the regression model used in Step 2 of Algorithm \ref{alg:drlearner}, trained instead using $\EIFa$, converges to $\tilde f$ at a fast enough rate (i.e., a parametric $\sqrt{n}$-rate), then the rate at which $\hat{\tilde f}$ converges to $\tilde f$ depends on the convergence rates of $\hat\g$, $\hat\m$, $\hat\rr$, and $\hat\bb$. Stated differently, if the conditions of Lemma \ref{th:rule1} are satisfied, then $\hat{\tilde f}$ will converge to $\tilde f^*$ asymptotically. This property is known as oracle efficiency; for a comprehensive discussion, see \cite{kennedy2023towards}.

\subsection{Estimating the expected counterfactual outcome value under the ODTR}\label{sec:ismo}

After estimating the ODTR, we now face the challenge of re-evaluating the rule’s performance in the pooled data; that is, determining how to apply the rule to the observations for which $V_2$ wasn't measured. While we can't directly estimate the CPE for these observations, we can estimate bounds. 

\begin{equation}\label{eq:bounds}
    \argmin_{v_2^*} \tilde\tau(\vone_{,i}, v_2^*)\leq \tilde\tau(\vone_{,i}, v_{2,i}) \leq \argmax_{v_2^*} \tilde\tau(\vone_{,i}, v_2^*)
\end{equation}

We have patients for whom the above bounds (\ref{eq:bounds}) give a clear decision rule and patients for whom it does not. If, for a patient $i$, the sign of $\argmin_{v_2^*} \hat{\tilde\tau}(\vone_{,i}, v_2^*)$ equals the sign of $\argmax_{v_2^*} \hat{\tilde\tau}(\vone_{,i}, v_2^*)$ the treatment rule would be the same regardless if we knew $v_{2,i}$; we refer to the treatment decision in this case as \textit{decisive}. For the set of patients where the sign of the bounds on the CPE are different, we refer to the treatment decision as \textit{ambiguous}.
Let $\d_1$ be the rule that treats all the patients for whom the rule is decisive according to the above bounds and also treats all the patients for whom the rule is ambiguous with $A=1$; conversely, let $\d_0$ be the rule that treats all the patients for whom the rule is decisive according to the above bounds but treats all patients for whom the rule is ambiguous with $A=0$.  We can then estimate the intervention-specific mean outcomes had these treatment decisions been applied---$\E[Y_{\d_1}]$ and $\E[Y_{\d_0}]$---using a variety of techniques from the causal inference literature. Under the CPE identifications assumptions \ref{ass:consistency}, \ref{ass:pos}, \ref{ass:noconfounding}, and that $\P(\tilde\tau(\vone,v_2) = 0) = 0$, these estimands are identified from the observed data \citep{van2015targeted}. For our application we choose to estimate $\E[Y_{\d_1}]$ and $\E[Y_{\d_0}]$ using targeted minimum-loss based estimation (TMLE) \citep{van2011targeted, van2015targeted}. We discuss TMLE further in \S\ref{sec:application}.

\subsection{Simulation study}

We performed a simulation study to evaluate the finite sample performance of the proposed estimator. We considered the following data-generating mechanism, where $\text{expit}(x) = (1 + e^{-x})^{-1}$ is the inverse logit function.
\begin{align*}
      \P(W_1 = 1) &= 0.33 \\
     W_2 &\sim \text{Beta}(2, 2) \\
     \P(V_{1,1} = 1\mid \W) &= \text{expit}(0.5 - 0.2W_1 + 0.15W_2) \\
     \P(V_{1,2} = 1\mid \W) &= \text{expit}(-0.3 + 0.1W_1 - 0.6W_2) \\
     \P(V_{1,3} = 1\mid \W) &= \text{expit}(0.1 + 0.3W_1 + 0.2W_2) \\
     \P(V_2 = 1\mid \Vone,\W) &= \text{expit}(-0.5 + 0.6V_{1,1} - 0.4V_{1,2} + 0.3V_{1,3} + 0.1W_1 - 0.2W_2)\\
     \P(S = 1 \mid \Vone, \W) &= \text{expit}(0.5W_1 - 0.3W_2 + 0.2V_{1,1} - 0.4V_{1,2} + 0.3V_{1,3}) \\ 
     \P(A = 1\mid S) &= 0.5S \\
     \P(Y = 1\mid A,\Vone,V_2,\W) &= \text{expit}(-1.5 + 0.3W_1 - 0.4W_2 + 0.1A + 0.5V_{1,1} - 0.8V_{1,2} + 0.2V_{1,3}\\
     &\hspace{4em} + V_{1,1}A - 1.2V_{1,2}A + 0.5V_{1,3}A + 0.9V_2 + 1.2V_2A)   
\end{align*}
The true values of $\tilde{\tau}(\Vone, V_2)$ and $\tau(\Vone, V_2)$ are provided in the supplementary materials (Table \ref{tab:truth}). For samples size $n \in (500, 1000, 2500, 10000)$ we conducted $K = 1000$ simulations and compared the performance of the proposed estimators to a plug-in estimator. All nuisance parameters were estimated using saturated cross-validated generalized-linear models (GLM) with $\ell1$-penalization (i.e., LASSO) \citep{tibshirani1996regression} and were calibrated using isotonic calibration \citep{van2024automatic}. The number of cross-fitting folds was chosen according to sample size: $n = 500$, 20-folds; $n \in (1000, 2500)$, 10-folds; $n = 10000$, 2-folds. Following \cite{kennedy2017non}, estimator performance was evaluated in terms of absolute mean bias and root-mean-squared error (RMSE), both summed over $\P(\Vone = \vone, V_2 = v_2)$. 

\begin{align*}
    \hat{\text{Bias}} &= \sum_{\mathcal{V}_1\times\mathcal{V}_2}\bigg| \frac{1}{K} \sum_{k=1}^K \hat{\tilde\tau}_k(\vone,v_2) - \tilde{\tau}_k(\vone,v_2)\bigg|\P(\vone,v_2)\\
    \hat{\text{RMSE}} &= \sum_{\mathcal{V}_1\times\mathcal{V}_2} \bigg[\frac{1}{K} \sum_{k=1}^K \big( \hat{\tilde\tau}_k(\vone,v_2) - \tilde{\tau}_k(\vone,v_2) \big)^2\bigg]^{-\frac{1}{2}}\P(\vone,v_2)
\end{align*}

The results of the simulation study are shown in Table \ref{tab:sim}. Across all samples sizes, the proposed estimator (DR-learner) had a smaller estimated bias than the plugin estimator. Only at the largest sample size (10000) did the DR-learner have a smaller estimated RMSE compared to the plugin estimator. 

\begin{table}[htbp]
\centering
\begin{threeparttable}
\caption{Comparison of estimated integrated mean absolute bias and root-mean-squared-error (RMSE) between the DR-learner and a plugin estimator across different sample sizes.}\label{tab:sim}
\begin{tabular}{rcccc}
\toprule
\multicolumn{1}{c}{ } & \multicolumn{2}{c}{DR-learner} & \multicolumn{2}{c}{Plugin} \\
\cmidrule(l{3pt}r{3pt}){2-3} \cmidrule(l{3pt}r{3pt}){4-5}
n & $\hat{\text{Bias}}$ & $\hat{\text{RMSE}}$ & $\hat{\text{Bias}}$ & $\hat{\text{RMSE}}$\\
\midrule
500  & 0.006 & 0.118 & 0.032 & 0.070\\
1000 & 0.003 & 0.081 & 0.032 & 0.055\\
2500 & 0.003 & 0.051 & 0.028 & 0.042\\
10000 & 0.002 & 0.025 & 0.025 & 0.031\\
\bottomrule
\end{tabular}
\end{threeparttable}
\end{table}

\section{Application to randomized trials of medications for opioid use disorder}

\subsection{Data and estimation details}\label{sec:application}

In an attempt to improve upon the dynamic treatment rule from \citet{rudolph2023optimally}, we applied the proposed estimator to the harmonized cohort of CTN0051 and CTN0030 to learn a dynamic treatment rule based on patient characteristics---including housing stability (i.e., $V_2$)---for assigning BUP-NX or XR-NTX to minimize the risk of relapse after 12-weeks of treatment. In CTN0051 ($n = 570$), patients were randomized to receive either BUP-NX or XR-NTX while in CTN0030 ($n = 360$) patients were randomized to receive BUP-NX or BUP-NX plus individual drug counseling. Based on \citet{rudolph2021optimizing} and \citet{rudolph2023optimally}, we included biological sex, age, history of alcohol use disorder, history of schizophrenia, history of anxiety or panic disorder, history of depression, past 30-day cannabis use, and past 30-day amphetamine use as baseline covariates (i.e., $\W$); we included race, severity of withdrawal symptoms, history of cocaine use disorder, history of IV drug use, history of epilepsy, and past 30-day use of benzodiazepines as the set of potential measured effect modifiers (i.e., $\Vone$).

We estimated nuisance parameters using an ensemble \citep{van2007super} of an intercept-only model, a main-effect generalized linear model, multivariate adaptive regression splines (MARS) \citep{friedman1991multivariate}, random forests \citep{wright2017ranger}, Bayesian additive regression trees (BART) \citep{chipman2010bart}, and LASSO \citep{tibshirani1996regression}. Nuisance parameters were cross-fit with 10-folds and calibrated using isotonic calibration \citep{van2024automatic}. The second-stage pseudo-outcome regressions were estimated using XGBoost \citep{chen2016xgboost} with the best hyperparameter specification chosen from a grid search using nested resampling \citep{bischl2023hyperparameter}. We evaluated feature importance and model explainability using SHapley Additive exPlanations (SHAP) values \citep{lundberg2017unified}. SHAP values are a model agnostic method for explaining predictions from a model. Specifically, a SHAP value can be interpreted as the change in a models prediction when conditioning on a specific variable. We estimated the expected risk of week-12 relapse if patients were assigned treatment according to the learned rule, a learned rule---using the CATE---that ignored housing status, a static rule where all patients receive BUP-NX, and a static rule where all patients receive XR-NTX using TMLE with cross-fitting (10-folds). TMLE is a doubly-robust substitution estimator that updates initial estimates of nuisance parameters to solve the estimating equation
\[n^{-1}\sum_{i=1}^n \hat\xi(\mathbf{O}_i, \hat\eta) = 0\]
where $\hat\xi(\mathbf{O}, \hat\eta)$ is the estimated EIF of the respective     parameter. We estimated the variance using the sample variance of the EIF. We refer the reader to \cite{gruber2009targeted} for further information on TMLE. All analyses were performed in \texttt{R} 4.4.2 \citep{Rbase} and our code is available on \href{https://github.com/CI-NYC/odtr-missing-modifiers}{GitHub}. 

\subsection{Results}

\begin{figure}[H]
  \centering
  \includegraphics[width=\textwidth]{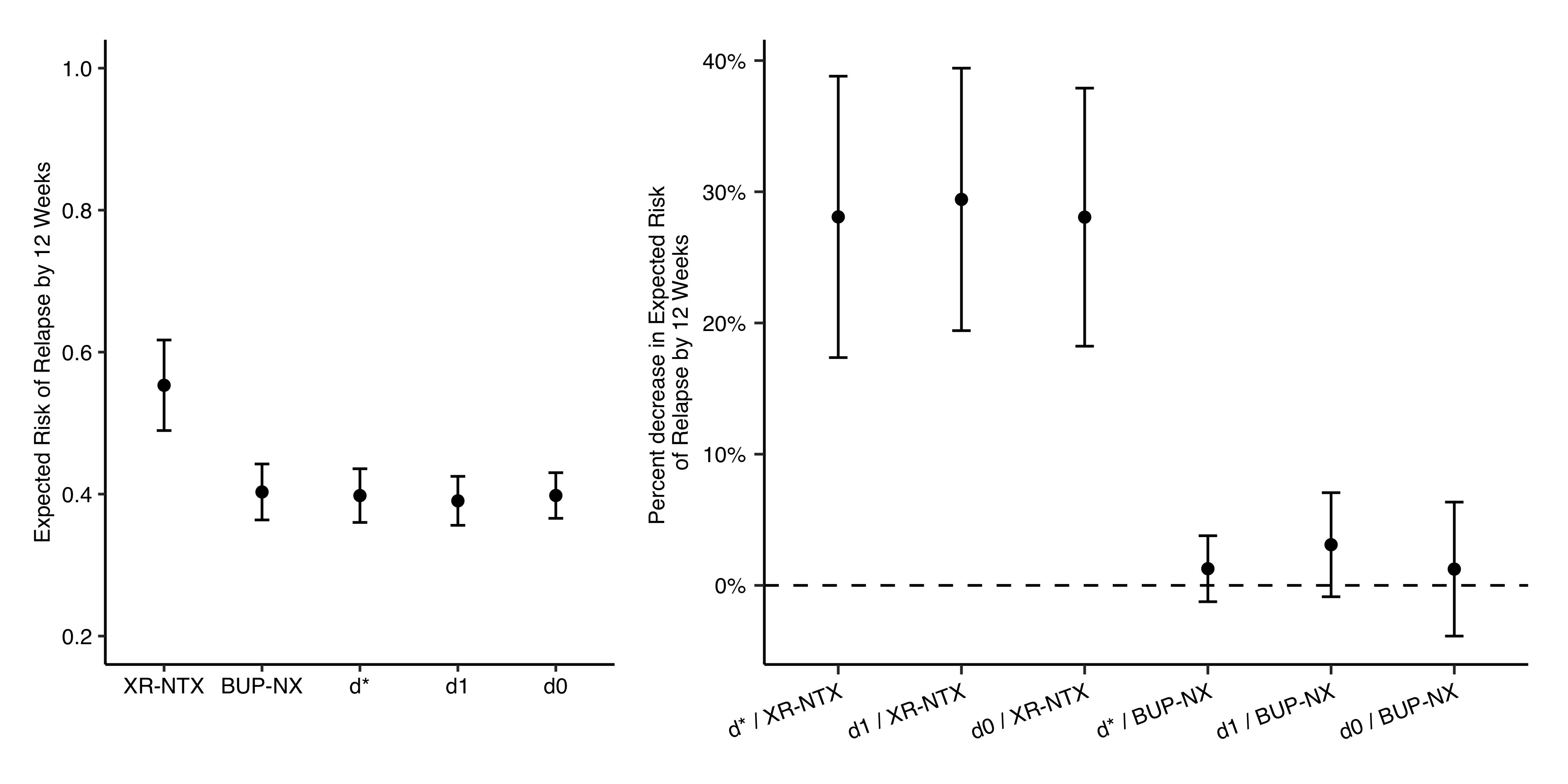}
  \caption{Estimated expected risk of week-12 relapse under various treatment rules (left panel) and estimated percent decrease in risk of week-12 relapse comparing learned treatment rules to static treatment rules (right panel). Treatment rules are as follows: XR-NTX, treat all patients with XR-NTX; BUP-NX, treat all patients with BUP-NX; $\d^*$, learned ODTR ignoring housing status; $d1$, learned ODTR using the proposed estimator and assigning patients with ambiguous treatment decisions BUP-NX; $\d0$, learned ODTR using the proposed estimator and assigning patients with ambiguous treatment decisions XR-NTX.}
  \label{fig:results}
\end{figure}

Results are shown in Figure \ref{fig:results}. Without knowing homeless status, 277 observations (78\%) from CTN0030 had decisive treatment decisions. We therefore estimated the expected risk of week-12 relapse under the learned treatment rule two-ways: assigning observations with an ambiguous decision BUP-NX ($\d_1$) or assigning observations with an ambiguous decision XR-NTX ($\d_0$). The estimated, expected risk of week-12 relapse under a static treatment rule where all patients were assigned XR-NTX was 0.55 (95\% CI: 0.49, 0.62); under a static rule where all patients were assigned BUP-NX was 0.4 (95\% CI: 0.36, 0.44); under the learned treatment rule that ignored housing status was 0.4 (95\% CI: 0.36, 0.44); under the learned treatment rule that assigned ambiguous patients BUP-NX was 0.39 (95\% CI: 0.36, 0.43); and, under the learned treatment rule that assigned ambiguous patients XR-NTX was 0.4 (95\% CI: 0.37, 0.43). We estimated that assigning treatment based on the learned rule that assigned ambiguous patients BUP-NX would decrease week-12 relapse by 29\% (95\% CI: 19\%, 39\%) compared to always treating with XR-NTX and 3\% (95\% CI: -1\%, 7\%) compared to always treating with BUP-NX. We also estimated that assigning treatment according to the rule that ignored housing status would decrease week-12 relapse by 28\% (95\% CI: 17\%, 39\%) compared to always treating with XR-NTX and 1\% (95\% CI: -1\%, 4\%) compared to always treating with BUP-NX. Comparatively, we found that the learned rule that incorporated housing status and assigned ambiguous patients BUP-NX would decrease week-12 relapse by 2\% (95\% CI: -2\%, 6\%) compared to the rule that ignored housing status. 

\begin{figure}[H]
  \centering
  \includegraphics[width=\textwidth]{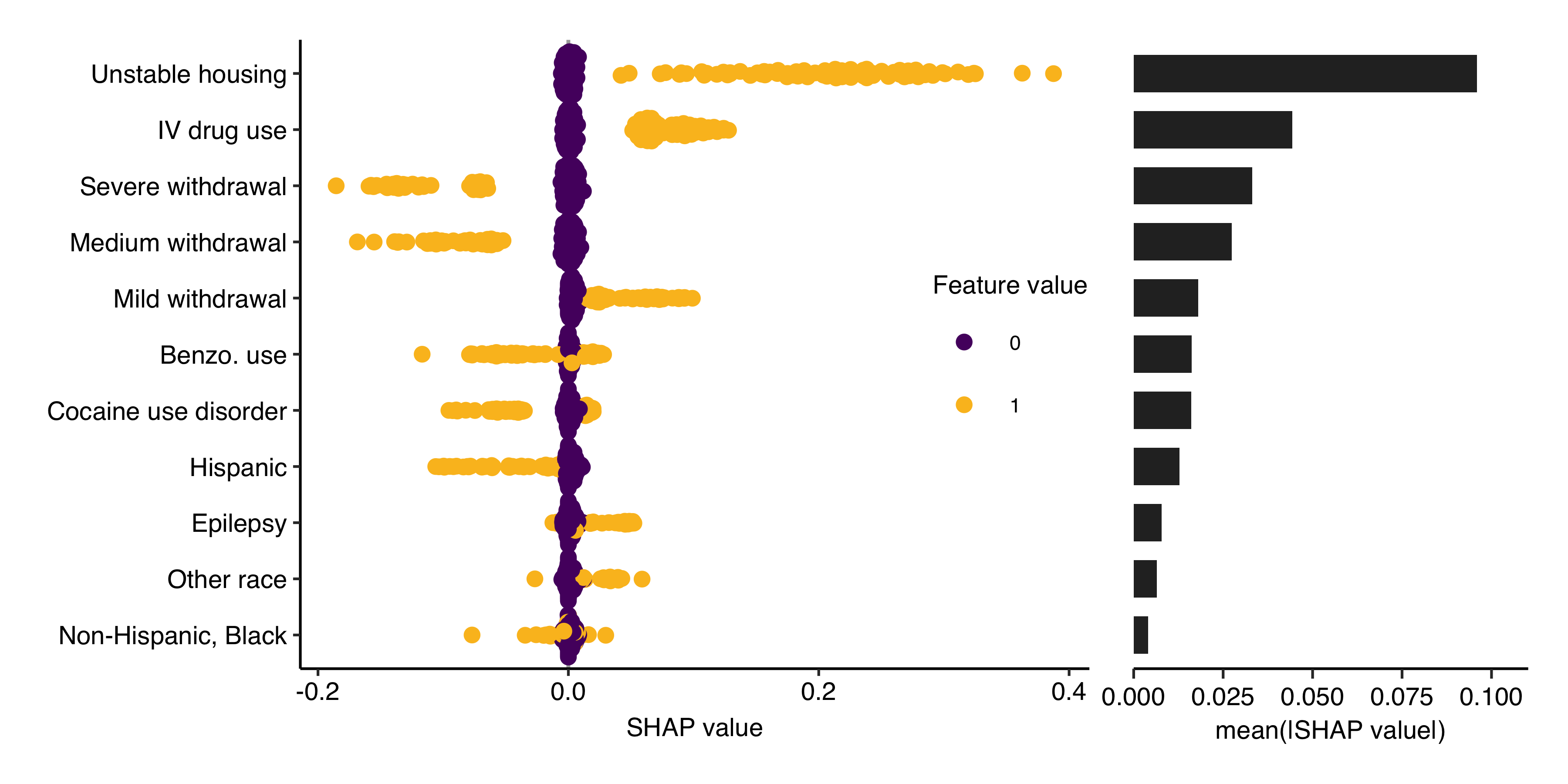}
  \caption{Beeswarm plot of SHAP values (left panel) and variable importance based on SHAP values (right panel).}
  \label{fig:shap}
\end{figure}

Based on the average absolute SHAP value, the variable with the largest influence on the CPE was housing status followed by history of IV drug use, severe withdrawal symptoms, and medium withdrawal symptoms (Figure \ref{fig:shap}). Generally, being unhoused, having a history of IV drug use, or experiencing only mild withdrawal symptoms increased the predicted CPE---indicating these patients may be better treated with XR-NTX---while experiencing severe or medium withdrawal symptoms decreased the predicted CATE---indicating these patients may be better treated with BUP-NX.

\section{Discussion}\label{sec:discussion}

In this paper, we explored the estimation of an ODTR using data from fused RCTs characterized by systematic missingness of a discrete effect modifier. Our approach is based on a novel decomposition of the CATE, which yields an alternative target estimand, the CPE, and requires weaker conditions for consistent estimation. We proposed a cross-fitted doubly robust estimator, the DR-Learner, for the CPE and, in a small simulation study, demonstrated that the proposed estimator outperforms a naïve plug-in estimator. Finally, we applied the DR-Learner to two fused comparative effectiveness RCTs of medication-assisted treatment for OUD.

\subsection{Implications of the identification assumptions}\label{sec:implications}

Learning an ODTR based on the CPE requires a set of untestable assumptions to identify the CPE from observed data. Consequently, the plausibility of these assumptions, and the extent to which they are likely to be satisfied in our motivating application, must be evaluated using substantive knowledge. Consistency (\ref{ass:consistency}), positivity (\ref{ass:pos}), and exchangeability (\ref{ass:noconfounding}) are standard in causal inference, and we expect them to be met given that the data were collected from RCTs. 

The unmeasured modifier treatment exchangeability assumption (\ref{ass:amod}), unmeasured modifier $S$-admissibility assumptions (\ref{ass:smod}), and joint positivity assumptions (\ref{ass:joint}) are unique to this work and will hold only under certain conditions. Namely, assumptions \ref{ass:amod} and \ref{ass:smod} will only hold when $V_2$ is a mediator on the causal pathway from $\W\backslash\Vone$ to $Y$ (i.e., $\W\backslash\Vone \rightarrow V_2 \rightarrow Y$). We believe this assumption is plausible in our data. Many of the variables we included in $\W$ and $\Vone$ cannot be caused by an individuals housing status (e.g., biological sex, age, race). The causal ordering of housing status and variables such as past drug-use, however, is more difficult to disentangle and is impossible to verify without the trials having collected more detailed information from patients. Assumption \ref{ass:amod} requires all the common causes of $V_2$ and $A$ be measured in $\W$, $\Vone$, and $Y_a$; again, because treatment was randomized within the trials we expect this assumption to be satisfied. As was briefly discussed in \S\ref{sec:identification}, \ref{ass:smod} is essentially a transport assumption. In the context of our analysis, it means that conditional on $\W$, $\Vone$, and $Y_a$, the probability of a trial participant being unhoused is the same in both trials.  

\subsection{Discussion of application results}\label{sec:applicationresults}

Using the proposed estimator, we derived an ODTR for assigning BUP-NX or XR-NTX to treat OUD using data from two RCTs, one of which did not measure participants' housing status. We then estimated the expected risk of week-12 relapse under treatment assignment based on the ODTR and compared it to the expected risk under static interventions and an ODTR that ignored housing status. Both of the learned rules significantly reduced the risk of week-12 relapse compared to always assigning XR-NTX. Relative to the rule that ignored housing status, we estimated a lower risk of week-12 relapse when applying the rule that incorporated housing status and assigned patients with ambiguous treatment decisions to BUP-NX; however, this difference was not statistically significant.

Using SHAP values to evaluate variable importance, we found that housing status contributed the most to the CPE, followed by IV drug use and withdrawal symptoms. The SHAP analysis also supported our hypothesis that patients with unstable housing may experience better outcomes with XR-NTX due to its administration protocol, while those with mild or severe withdrawal symptoms may benefit more from BUP-NX, given the induction requirements for XR-NTX. Ultimately, our results provide further evidence that clinicians should consider patients’ housing status when selecting a course of medication-assisted treatment for OUD.

\begin{funding}
This work was supported by the National Institute of Drug Abuse (R01DA056407).
\end{funding}


\bibliographystyle{imsart-nameyear} 
\bibliography{lib}       

\supplement

\begin{table}[H]
\centering
\begin{threeparttable}
\caption{True values of the CPE ($\tilde{\tau}(\vone, v_2)$) and the CATE ($\tau(\vone, v_2)$) for the data-generating mechanism used in the simulation study.}\label{tab:truth}
\begin{tabular}[t]{ccccccc}
\toprule
$V_{1,1}$ & $V_{1,2}$ & $V_{2,1}$ & $V_{2,2}$ & $\P(\vone, v_2)$ & $\tau(\vone, v_2)$\tnote{a} & $\tilde{\tau}(\vone, v_2)$\tnote{b}\\
\midrule
0 & 1 & 1 & 0 & 0.052 & -0.044 & -0.029\\
1 & 0 & 1 & 1 & 0.133 & 0.443 & 0.257\\
1 & 1 & 0 & 1 & 0.040 & 0.252 & 0.103\\
1 & 0 & 0 & 1 & 0.086 & 0.443 & 0.224\\
1 & 1 & 0 & 0 & 0.057 & -0.011 & -0.006\\
0 & 0 & 1 & 0 & 0.078 & 0.114 & 0.064\\
1 & 0 & 0 & 0 & 0.084 & 0.247 & 0.122\\
1 & 1 & 1 & 1 & 0.063 & 0.378 & 0.183\\
0 & 1 & 0 & 1 & 0.016 & 0.017 & 0.005\\
0 & 1 & 0 & 0 & 0.042 & -0.055 & -0.040\\
0 & 0 & 1 & 1 & 0.060 & 0.406 & 0.176\\
1 & 0 & 1 & 0 & 0.096 & 0.376 & 0.158\\
0 & 0 & 0 & 1 & 0.037 & 0.310 & 0.112\\
0 & 1 & 1 & 1 & 0.027 & 0.118 & 0.040\\
1 & 1 & 1 & 0 & 0.067 & 0.060 & 0.031\\
0 & 0 & 0 & 0 & 0.065 & 0.014 & 0.009\\
\bottomrule
\end{tabular}
\begin{tablenotes}
\footnotesize
\item[a] Conditional average treatment effect (CATE).
\item[b] Conditonal proxy effect (CPE). 
\end{tablenotes}
\end{threeparttable}
\end{table}

\subsection{Conditional mean specific outcome decomposition}

\begin{proof}\label{proof:cmsd}
    \begin{align*}
    \P(Y_a = 1\mid \Vone, V_2)
    &= \frac{\P(Y_a=1, \Vone, V_2)}{\P(\Vone, V_2)}\\
    &= \frac{\P(V_2\mid Y_a =1, \Vone)\P(Y_a=1,\Vone)}{\P(V_2\mid \Vone)\P(\Vone)}\\
    &= \frac{\P(V_2\mid Y_a =1, \Vone)\P(Y_a=1\mid \Vone)\P(\Vone)}{\P(V_2\mid \Vone)\P(\Vone)}\\
    &= \frac{\begingroup\color{red}\P(V_2\mid Y_a =1, \Vone) \endgroup\begingroup\color{blue}\P(Y_a=1\mid \Vone)\endgroup}{\begingroup\color{red}\P(V_2\mid \Vone)\endgroup}.
\end{align*}
The red densities only be estimated among observations where $S=1$, while the blue density can be estimated among all observations.
\end{proof}

\subsection{Proof of Lemma \ref{th:identification}}

\begin{proof}
Under assumptions \ref{ass:consistency}-\ref{ass:joint}, $\tilde f(a, \vone, v_2) = \P(V_2=v_2\mid Y_a=1, \Vone)\P(Y_a=1\mid \Vone=\vone)$ is identified from the observed data as
\begin{equation*}
    \int \bb(a,v_2,\vone)\m(a,\vone)\P(\mathbf{w}\mid\vone)\,d\mathbf{w},
\end{equation*}
where 
\begin{align*}
\bb(a,v_2,\vone) &=\P(V_2=v_2\mid Y=1,A=a,\W,\Vone=\vone,S=1) \\
\m(a,\vone) &= \E[Y\mid A=a,\Vone=\vone,\W].
\end{align*}
We first show that  $\P(Y_a=1\mid \Vone) = \E[\E[Y\mid A=a, \W,\Vone]\mid \Vone]$. 
\begin{align*}
\P(Y_a=1\mid \Vone)&=\E[Y_a\mid \Vone]\\
&= \E[\E[Y_a\mid \W,\Vone]\mid \Vone]\\
&= \E[\E[Y\mid A=a,\W,\Vone]\mid \Vone]
\end{align*}
The second equality follows from the tower rule; the third from consistency (\ref{ass:consistency}) and exchangeability (\ref{ass:noconfounding}). We can identify $\P(V_2\mid Y_a=1,\Vone)$ by: 
\begin{align*}
        \P(V_2=v_2\mid Y_a=1,\Vone) &= \E[\P(V_2=v_2\mid Y_a=1,\Vone,\W)\mid Y_a=1,\Vone]\\
        &= \int \P(V_2=v_2\mid Y_a=1,\Vone,\W)\P(\mathbf{w}\mid Y_a=1,\Vone) \,d\mathbf{w}\\
        &= \int \P(V_2=v_2\mid Y_a=1,A=a,\Vone,\W)\P(\mathbf{w}\mid Y_a=1,\Vone) \,d\mathbf{w}\\
        &= \int \P(V_2=v_2\mid Y=1,A=a,\Vone,\W)\P(\mathbf{w}\mid Y_a=1,\Vone) \,d\mathbf{w}
\end{align*}
The first and second equalities follow from the tower rule and the definition of conditional expectation; the third from unmeasured modifier treatment exchangeability (\ref{ass:amod}), and the fourth from consistency (\ref{ass:consistency}). $\P(\W\mid Y_a=1,\Vone)$ is identified as follows. 
\begin{align*}
    \P(\W\mid Y_a=1,\Vone) &= \frac{\P(\W,Y_a=1,\Vone)}{\P(Y_a=1,\Vone)} \\&= \frac{\P(Y_a=1\mid \Vone,\mathbf{w})\P(\W\mid \Vone)\P(\Vone)}{\P(Y_a=1\mid \Vone)\P(\vone)} \\&= \frac{\P(Y_a=1\mid \Vone,\W)\P(\W\mid \Vone)}{\P(Y_a=1\mid \Vone)}\\
    &= \frac{\E[Y\mid A=a,\Vone,\W]\P(\W\mid \Vone)}{\P(Y_a=1\mid \Vone)}
\end{align*}
The fourth equality follows from consistency (\ref{ass:consistency}) and exchangeability (\ref{ass:noconfounding}). Plugging this back in yields 
\begin{align*}
    \P(V_2\mid Y_a=1,\Vone) &= \int \frac{\P(V_2\mid Y=1,A=a,\Vone,\W)\E[Y\mid A=a,\Vone,\mathbf{w}]\P(\mathbf{w}\mid \Vone)}{\P(Y_a=1\mid \Vone)} \,d\mathbf{w} \\
    &= \frac{\E[\P(V_2\mid Y=1,A=a,\Vone,\W)\E[Y\mid A=a,\Vone,\W]\mid \Vone]}{\E[\E[Y\mid A=a,\W,\Vone]\mid \Vone]}\\
    &= \frac{\E[\P(V_2\mid Y=1,A=a,\Vone,\W,S=1)\E[Y\mid A=a,\Vone,\W]\mid \Vone]}{\E[\E[Y\mid A=a,\W,\Vone]\mid \Vone]}
\end{align*}
The second equality uses the definition of conditional expectation and the identification result for $\P(Y_a=1\mid \Vone)$. The third equality follows from unmeasured modifier $S$-admissibility (\ref{ass:smod}). Plugging in the identification results for $P(Y_a=1\mid V_1)$ and $\P(V_2\mid Y_a=1,\Vone)$ yields
\begin{align*}
    \E[\P(V_2\mid Y=1,A=a,\Vone,\W,S=1)\P(Y=1\mid A=a,\Vone,\W)\mid \Vone],
\end{align*}
which is equivalent to
\[\int \bb(a,v_2,\vone)\m(a,\vone)\P(\mathbf{w}\mid\vone)\,d\mathbf{w}.\]

\end{proof}

\subsection{Efficient influence function of $\kappa(a,v_2)$}

\begin{proof}
To derive the EIF of $\kappa(a,v_2)$ we assume that $\W$ and $\Vone$ are discrete so that integrals can be treated as sums. Then, 
\begin{align*}
    \kappa(a,v_2) = \sum_{\mathcal{W}\times\mathcal{V}_1} \bb(a,v_2,\vone)\m(a,\vone)\P(\w,\vone)\text{.}
\end{align*}
We define $\phi [\psi]$ as the operator that returns the EIF of a statistical parameter $\psi$. We then have
\begin{align*}
\phi[\kappa(a,v_2)] &= \phi \bigg[\sum_{\mathcal{W}\times\mathcal{V}_1}\bb(a,v_2,\vone)\m(a,\vone)\P(\w,\vone) \bigg] \\
&= \sum_{\mathcal{W}\times\mathcal{V}_1} \phi \big[\bb(a,v_2,\vone)\m(a,\vone)\P(\w,\vone) \big] \\
&= \underbrace{\sum_{\mathcal{W}\times\mathcal{V}_1}\m(a,\vone)\P(\w,\vone)\phi[\bb(a,v_2,\vone)]}_{\varphi_\bb(a,v_2)} + \\ &\quad\underbrace{\sum_{\mathcal{W}\times\mathcal{V}_1}\bb(a,v_2,\vone)\P(\w,\vone)\phi[\m(a,\vone)]}_{\varphi_\m(a,v_2)} + \\
&\quad\underbrace{\sum_{\mathcal{W}\times\mathcal{V}_1}\bb(a,v_2,\vone)\m(a,\vone)\phi[\P(\w,\vone)]}_{\varphi_\kappa(a,v_2)} 
\end{align*}
where the third equality follows from treating EIFs as derivatives and applying the product rule: $\phi[f(x)g(x)] = g(x)\phi[f(x)] + f(x)\phi[g(x)]$. The EIFs of a conditional expectation and a conditional probability mass function are well established. First, define the nuisance parameters 
\begin{align*}
    \g(a,\vone) &= \P(A=a\mid\Vone=\vone,\W)\\
    \rr(a,\vone) &= \P(Y=1,A=a,S=1\mid \Vone=\vone,\W),
\end{align*}
and the weights
\[
\begin{alignedat}{2}
    \alpha_\g(a,\vone) &= \frac{\one(A=a)}{\g(a,\vone)} 
    &\quad \alpha_{\rr}(a,\vone) &= \frac{\one(A=a,Y=1,S=1)}{\rr(a,\vone)}.
\end{alignedat}
\] 
The EIFs of $\m(a,\vone)$, $\bb(a,v_2,\vone)$, and $\P(\w,\vone)$ are then
\begin{align*}
&\phi[\bb(a,v_2,\vone)] = \alpha_\rr(a,\Vone)\big\{\one(V_2=v_2) - \bb(a,v_2,\Vone)\big\}\\
&\phi[\m(a,\vone)] = \alpha_\g(a,\Vone)\big\{Y - \m(a,\Vone)\big\}\\
&\phi[\P(\w,\vone) = \one(\W=\w,\Vone=\vone) - \P(\w,\vone). 
\end{align*}
We can then plug these values back in which yields the components
\begin{align*}
\varphi_\bb(a,v_2) &= \sum_{\mathcal{W}\times\mathcal{V}_1}\m(a,\vone)\P(\w,\vone)\phi[\bb(a,v_2,\vone)] \\
    &= \m(a,\Vone)\phi[\bb(a,v_2,\Vone)] \\
    &= \alpha_\rr(a,\Vone)\big\{\one(V_2=v_2)\m(a,\Vone)- \bb(a,v_2,\Vone)\m(a,\Vone)\big\},
\end{align*}
and
\begin{align*}
\varphi_\m(a,v_2) &= 
    \sum_{\mathcal{W}\times\mathcal{V}_1}\bb(a,v_2,\vone)\P(\w,\vone)\phi[\m(a,\vone)]\\
    &= \bb(a,v_2,\Vone)\phi[\m(a,\Vone)] \\
    &= \alpha_\g(a,\Vone)\big\{Y\bb(a,v_2,\Vone) - \bb(a,v_2,\Vone)\m(a,\Vone)\big\},
\end{align*}
and
\begin{align*}
\varphi_\kappa(a,v_2) &= 
    \sum_{\mathcal{W}\times\mathcal{V}_1}\bb(a,v_2,\vone)\m(a,\vone)\phi[\P(\w,\vone)] \\
    &= \sum_{\mathcal{W}\times\mathcal{V}_1}\bb(a,v_2,\vone)\m(a,\vone)\big\{\one(\W=\w,\Vone=\vone) - \P(\w,\vone)\big\} \\
    &= \bb(a,v_2,\Vone)\m(a,\Vone) - \kappa(a,v_2).
\end{align*}
Recombining terms yields
\[\varphi_\bb(a,v_2) + \varphi_\m(a,v_2) + \bb(a,v_2,\Vone)\m(a,\Vone) - \kappa(a,v_2)\]
where $\varphi_\bb(a,v_2) + \varphi_\m(a,v_2) + \bb(a,v_2,\Vone)\m(a,\Vone)$ is the uncentered efficient influence function of $\kappa(a,v_2)$ and is the result in the main text.
\end{proof}

\subsection{Proof of Lemma \ref{th:remainder}}

\begin{proof}
\begin{align*}
    \text{Rem}(a, \vone, v_2;\eta') &= \tilde f_P - \E_P[\xi(O;\eta')\mid\Vone=\vone]\\
&= \tilde f_P - \E_P\big[\frac{\one(A=a,Y=1,S=1)}{\rr'}\cdot \m' \cdot\big\{\one(V_2=v_2) - \bb' \big\} \mid \Vone=\vone\big] \\
&\quad - \E_P\big[\frac{\one(A =a)}{\g'} \cdot \bb' \cdot \big\{Y - \m' \big\} \mid \Vone=\vone\big] \\
&\quad - \E_P\big[\bb' \cdot \m' \mid \Vone = \vone\big] \\
&= \underbrace{- \E_P\big[\frac{\one(A=a,Y=1,S=1)}{\rr'} \cdot \m' \cdot \big\{\one(V_2=v_2) - \bb' \big\} \mid \Vone=\vone\big]}_{C'_1} \\
&\quad \underbrace{- \E_P\big[\frac{\one(A =a)}{\g'} \cdot \bb' \cdot \big\{Y - \m' \big\} \mid \Vone=\vone\big]}_{C'_2} \\
&\quad + \underbrace{\E_P[\bb \cdot \m \mid \Vone=\vone] - \E_P\big[\bb' \cdot \m' \mid \Vone = \vone\big]}_{C'_3}\\
\end{align*}
The third equality follows from $\tilde f_P = \E_P[\bb\cdot\m \mid \Vone=\vone]$ and rearranging.

\begin{align*}C'_1 
&= - \E_P\big[\frac{\one(A=a,Y=1,S=1)}{\rr'} \cdot \m' \cdot \big\{\one(V_2=v_2) - \bb' \big\} \mid \Vone=\vone\big] \\
&= -\E_P\big[\m' \cdot \frac{\rr}{\rr'} \cdot \big\{\bb - \bb' \big\} \mid \Vone=\vone \big]\\
&= -\E_P\big[\m' \cdot \frac{\rr}{\rr'}\cdot\big\{\bb - \bb' \big\} + \m' \cdot \bb - \m'\cdot\frac{\rr'}{\rr'}\cdot\bb + \m'\cdot\frac{\rr'}{\rr'}\cdot\bb' - \m'\cdot\bb'  \mid \Vone=\vone \big]\\
&= -\E_P\big[\m'\cdot\frac{\rr}{\rr'}\cdot\big\{\bb - \bb' \big\} + \m'\cdot\bb - \m'\cdot\bb' - \m'\cdot\frac{\rr'}{\rr'}\cdot\bb + \m'\cdot\frac{\rr'}{\rr'}\cdot\bb' \mid \Vone=\vone \big]\\
&= -\E_P\big[\m'\cdot\frac{\rr}{\rr'}\cdot\big\{\bb - \bb' \big\} + \m'\cdot\bb - \m'\cdot\bb' - \m'\cdot\frac{\rr'}{\rr'}\cdot(\bb -\bb') \mid \Vone=\vone \big]\\
&= \underbrace{-\E_P\big[\m'\cdot\frac{(\rr - \rr')}{\rr'}\cdot(\bb - \bb')\mid\Vone=\vone\big]}_{C'_\bb} + \underbrace{\E_P\big[\m'\cdot\bb - \m'\cdot\bb' \mid \Vone=\vone\big]}_{C'^*}
\end{align*}
The second equality follows from the tower rule, the third from adding and subtracting 1 (twice), and the rest from rearranging.

\begin{align*}
    C'_2 &= - \E_P\big[\frac{\one(A =a)}{\g'}\cdot\bb'\cdot\big\{Y - \m' \big\} \mid \Vone=\vone\big] \\
    &=- \E_P\big[\frac{\g}{\g'}\cdot\bb'\cdot\big\{\m - \m' \big\} \mid \Vone=\vone\big] \\
    &= \underbrace{-\E_P\big[\bb'\cdot\frac{(\g - \g')}{\g'}\cdot\big\{\m - \m' \big\}\mid\Vone=\vone\big]}_{C'_\m} + \underbrace{\E_P\big[\bb'\cdot\m - \bb'\cdot\m' \mid \Vone=\vone\big]}_{C'^\#}
\end{align*}
The second equality follows from the tower rule, and the third from adding and subtracting 1 (twice) and rearranging using the same steps as above. Define $C'_\kappa = C'_3 + C'^* + C'^\#$. We then have
\[
    \text{Rem}(a, \vone, v_2;\eta') = C'_\bb + C'_\m + C'_\kappa\text{.}
\]
\begin{align*}
    C'_\kappa &= \E_P[\bb\cdot\m \mid \Vone=\vone] - \E_P[\hat\bb\cdot\hat\m \mid \Vone=\vone] + C'^* + C'^\#\\
    &= \E_P\big[\bb\cdot\m -  \hat\bb\cdot\hat\m \mid \Vone=\vone\big] + \E_P\big[\hat\m\cdot\bb - \hat\m\cdot\hat\bb \mid \Vone=\vone\big] + \E_P\big[\hat\bb\cdot\m - \hat\bb\cdot\hat\m \mid \Vone=\vone\big]\\
    &= \E_P\big[\bb\cdot\m - \hat\bb\cdot\hat\m - \bb\cdot\hat\m + \hat\bb\cdot\hat\m - \hat\bb\cdot\m + \hat\bb\cdot\hat\m\mid \Vone=\vone \big]\\
    &= \E_P\big[\bb\cdot\m - \bb\cdot\hat\m - \hat\bb\cdot\m + \hat\bb\cdot\hat\m \mid \Vone=\vone \big]\\
    &= \E_P\big[\bb\cdot(\m - \hat\m) - \hat\bb\cdot(\m - \hat\m)\mid \Vone=\vone\big]\\
    &= \E_P\big[(\bb - \hat\bb)(\m - \hat\m)\mid \Vone=\vone\big]
\end{align*}
Inspection of $C'_\bb$, $C'_\m$ and $C'_\kappa$ reveals that each of these terms will equal zero (and therefore also $\text{Rem}(a, \vone, v_2;\eta')$) if: 1) $\bb' = \bb$ and $\g' = \g$, or 2) $\m' = \m$ and $\rr' = \rr$, or 3) $\bb' = \bb$ and $\m' = \m$.
\end{proof}

\end{document}